\documentclass[prb,twocolumn,amsmath,amssymb,floatfix,showpacs]{revtex4}
\usepackage{graphicx}
\usepackage{bm}
\usepackage{amssymb}
\usepackage{color}
\usepackage{epsfig}
\usepackage{subfigure}
\begin{document}

\title{Delayed currents and interaction effects in mesoscopic capacitors\\}

\author{Zohar Ringel}
\email{zohar.ringel@weizmann.ac.il}

\affiliation{Department of Condensed Matter Physics,  Weizmann
Institute of Science, Rehovot 76100, Israel}

\author{O. Entin-Wohlman}

\altaffiliation{School of Physics and Astronomy, Beverly and
Raymond Sackler Faculty of Exact Sciences, Tel Aviv University,
Tel Aviv 69978, Israel}

\affiliation{Department of Physics, Ben Gurion University, Beer
Sheva 84105, Israel}

\affiliation{Albert Einstein Minerva Center for Theoretical
Physics, Weizmann Institute of Science, Rehovot 76100, Israel}

\author{Y. Imry}

\affiliation{Department of Condensed Matter Physics,  Weizmann
Institute of Science, Rehovot 76100, Israel}

\date{\today}

\begin{abstract}
We propose an alternative derivation for the dynamic admittance of
a gated quantum dot connected by a single-channel lead to an
electron reservoir.  Our derivation, which reproduces the result
of Pr\^{e}tre, Thomas, and B\"{u}ttiker for the universal charge-relaxation resistance, shows that at low
frequencies, the current leaving the dot lags after the entering one by the Wigner-Smith delay time. We compute the capacitance when
interactions are taken into account only on the dot within the
Hartree-Fock approximation and study the Coulomb-blockade
oscillations as a function of the Fermi energy in the reservoir. In
particular we find that those oscillations disappear when the dot
is fully `open', thus we reconcile apparently conflicting
previous results.

\end{abstract}



\maketitle

\section{Introduction}

\begin{figure*}
\includegraphics[width=12cm]{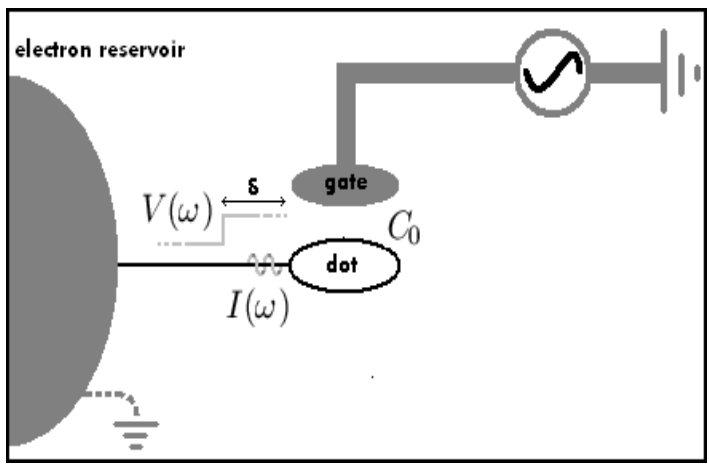}
\caption{The mesoscopic capacitor: the ac source excites a
periodic accumulation of charges on the gate and the latter
affects the charges on the dot via the potential $V$ created
in-between the dot and the gate. This in turn causes the ac
current $I$ flowing between the dot and the electron reservoir.
 } \label{SYS}
\end{figure*}

The low-frequency conductance of a mesoscopic conductor with capacitive
elements has recently attracted renewed interest due to the experimental
verification \cite{GL} of the fascinating prediction made in Ref.
~\onlinecite{B00} (see also Refs. ~\onlinecite{B0}, ~\onlinecite{B1} and ~\onlinecite{ROSA}), concerning the universal
value of the charge-relaxation resistance. The setup of the device  is
depicted in Fig. \ref{SYS}: a mesoscopic system consisting of a quantum dot
coupled by a single-channel wire to an electron reservoir. A macroscopic
gate, which is connected to an ac source, is placed nearby. This source
induces an ac potential on the dot, $U(\omega)$, and consequently a current,
$I(\omega )$, is flowing between the dot and the reservoir, such that
$I(\omega )=g(\omega )U(\omega )$, where $g(\omega )$ is the
frequency-dependent conductance of the dot. Note that in this formalism the
current $I$ is created by the {\em effective} potential on the dot (denoted
here by $U$) that includes screening effects,  and not by the bare potential,
$V$. As is shown in Ref. ~\onlinecite{B1}, the total ac conductance (often
referred to as admittance) of this device is $[g^{-1}_{\rm gate}(\omega
)+g^{-1}(\omega )+(-i\omega C_{0})^{-1}]^{-1}$, where $C_{0}$ is the
capacitance of the dot and the nearby gate, and $g_{\rm gate}(\omega )$ is
the conductance between the gate and the wire connecting it to the voltage
source. However, for large enough $C_{0}$ and $g^{}_{\rm gate}(\omega )$, the
ac conductance of the device, which is the quantity probed in the experiment,
is given by $g(\omega )$.

The ac conductance of noninteracting electrons can be presented in
terms of the scattering matrix of the mesoscopic system (see
Refs.~ \onlinecite{B1}, ~\onlinecite{BUTTI}, and
~\onlinecite{YEHOSHUA}). When the dot is connected by a
single-channel lead to an electron reservoir, as in Fig.
\ref{SYS}, the ac conductance is
\begin{align}
g(\omega) = \frac{e^2}{h}\int_{E_F-\hbar\omega}^{E_F} \frac{dE
}{\hbar\omega } \Bigl (1 - S^{\ast} (E)S(E+\hbar \omega)\Bigr )\ ,
\label{Free-G}
\end{align}
where
the scattering matrix $S(E)$ is just a phase factor
\begin{align}
S(E)=\exp [i\phi(E)]\ .\label{S-Phase}
\end{align}
In writing down Eq. (\ref{Free-G}) we have assumed low enough
temperatures, such that the Fermi functions factor
$f(E)-f(E+\hbar\omega)$ is replaced by unity within the indicated integration limits; $E_{F}$ is the Fermi energy of the reservoir.
The universal value of the charge-relaxation resistance predicted
in Ref.~ \onlinecite{B1}  and confirmed experimentally \cite{GL},
emerges upon comparing the low-frequency expansion of Eq.
(\ref{Free-G}) with the ac conductance, $g_{a}(\omega )$,  of a
conventional capacitor, $C$, connected  in series to a
dc-resistance, $R$,
\begin{align}
g_{a}(\omega )=-i\omega C+\omega^{2}C^{2}R+{\cal O}(\omega ^{3})\
.\label{RC}
\end{align}
One then finds for $R$ the value $h/2e^{2}$, which is half the
quantum unit of the resistance, and in particular is independent
of the scattering properties of the quantum dot, embedded in the
scattering matrix $S(E)$. The capacitance, on the other hand, is
given by $C=(e^{2}/2\pi )\phi '(E_F)$, where $\phi '(E_F)$ is the
energy-derivative of the phase (\ref{S-Phase}) at the Fermi energy.
Since it is related to coherent properties of the device, this capacitance is usually referred to as the ``quantum capacitance" \cite{B1,GL}.

Performing ac measurements without contacts or with only one ohmic contact might be very useful for molecules due to the difficulty in properly and nondestructively contacting them. However, the interest there is  in specific, more than in universal, properties. The quantum capacitance of a molecule may well therefore be of genuine interest.

The quantum capacitance is
expected to exhibit mesoscopic oscillations, related to Coulomb
blockade (CB), as function of the Fermi energy. \cite{BS} The
discussion of this effect necessitates the consideration of
electronic correlations in the system, notably on the quantum dot.
The original derivation of the low-frequency conductance used a Hartree-type approximation to capture the effects
of electronic interactions. \cite{B1} This was followed up in Ref.
~\onlinecite{ROSA}, and studied in great detail in particular in
Ref. ~\onlinecite{BS}. However, the latter reference reports on the
persistence of those oscillations even when the dot is fully open
to its lead (see Fig. 2 and Sec. V of Ref. ~\onlinecite{BS}). This
result has been obtained for relatively small values of
$E_{c}/{\rm d}$, where $E_{c}$ is the charging energy of the dot
and ${\rm d}$ is the mean-level spacing at the Fermi level.
Nonetheless, one should note that in the other limit, $E_{c}\gg
{\rm d}$,  Matveev\cite{MT} has shown that these oscillations
should disappear for a sufficiently large dot which is
completely open.

In this paper we expound on the two components of the low-frequency
conductance, the charge-relaxation resistance, $R$, and the quantum
capacitance, $C$. In the next section we present an alternative derivation of
the frequency-dependent  conductance (\ref{Free-G}), which emphasizes the
different roles played by the current flowing into the dot and the one
leaving it. In particular, our derivation sheds some light on the origin of
the intriguing observation of a universal-valued charge-relaxation
resistance, describing it in terms of a delayed current. Our simple model
does not include dephasing effects. For a clear discussion of those see Ref.
~\onlinecite{DEPHASING}. In Sec. \ref{CB} we analyze the effects of the
Coulomb interaction on the dot. We treat the electronic interactions within a
full Hartree-Fock calculation. In particular, we examine in Sec. \ref{CB} the
result of Ref. ~\onlinecite{BS} concerning the persistence of mesoscopic
oscillations in the capacitance when the dot is completely open, and show
that by including systematically all Fock correlations, the results are
reconciled with the prediction of Matveev \cite{MT}. Namely, we find that in
a completely open dot with a large number of equally spaced levels, the
on-dot interactions do not induce CB effects in the capacitance. Compared to
Matveev \cite{MT}, our treatment is not confined to strong interactions only.
It is, however, limited by our use of the mean-field approach. Moreover, our
theory ignores effects related to disorder which may modify the results
considerably (see for example Ref. ~\onlinecite{CHAOTIC}).

\section{Delayed currents}
\label{DC}

Here we show that the total current $I(t)$ (see Fig. \ref{SYS})
can be written in the form
\begin{align}
I(t)=\frac{e^{2}}{h}\Bigl (U(t)-U(t-\tau )\Bigr )\ ,\label{IT}
\end{align}
where $\tau$ is the Wigner-Smith \cite{SM} delay time,
\begin{align}
\tau =\hbar\frac{\partial\phi(E)}{\partial E}\ ,\label{WST}
\end{align}
evaluated at the Fermi energy \cite{COM2}. This result applies to the low-frequency ($\omega << 1/\tau$) components of $U(t)$ or to a slowly varying $U(t)$.
The result (\ref{IT}) gives a vivid interpretation of the dynamics of the device. The
current flowing between the dot and the reservoir consists of an
ingoing (into the dot) component, $I^{}_{\rm in}=(e^{2}/h)U(t)$,
which responds instantaneously to the effective voltage $eU(t)$, and an
outgoing one (away from the dot), $I^{}_{\rm
out}=(e^{2}/h)U(t-\tau )$, which lags after the effective voltage. The
incoming current is excited around the region where the potential
drop occurs  (see Fig. \ref{SYS}) and flows towards the dot; it
obeys the Landauer formula for an ideal single-channel wire. After
a  delay time $\tau$ that current  is reflected back and  flows
towards the reservoir as $I_{\rm out}$, again obeying the Landauer
formula for a perfect conductor. This means that in terms of the
Wigner-Smith delay time $\tau$, the conductance $g(\omega )$ is
\begin{align}
g(\omega )=\frac{e^{2}}{h}(1-e^{i\omega\tau})\ .\label{GTAU}
\end{align}
Comparing the low-frequency expansion of $g(\omega )$ of Eq.
(\ref{GTAU}) with the classical ac conductance of the equivalent RC
circuit, Eq. (\ref{RC}),  yields for the charge-relaxation
resistance, $R$, and the capacitor, $C$, the expressions \cite{B1}
\begin{align}
R=\frac{h}{2e^{2}}\ ,\ \  C=\frac{e^{2}}{h}\tau\ .\label{QUANT}
\end{align}
Note that $\tau = 2RC$. The proof of Eq. (\ref{IT}), or
alternatively, Eqs. (\ref{GTAU}) and (\ref{QUANT}), is based on
the linear-response expression for the ac conductance in the
framework of the scattering formalism, as derived e.g., in Refs.
~\onlinecite{B1}, ~\onlinecite{BUTTI}, and ~\onlinecite{YEHOSHUA}.
Our derivation (unlike  previous treatments employing the
scattering formalism) makes use of the {\em spatial dependence} of
the current operator in order to identify separately
 the incoming and the reflected
current {\em operators}. This separation, in turn, allows for the
calculation of the respective partial response functions of the two
currents. As a result, one finds that the incoming current obeys
the Landauer formula for a perfect (single-channel) wire, while
the reflected one is delayed by the Wigner-Smith time, leading
finally  to Eq. (\ref{IT}). This provides a qualitative
interpretation of the results of Ref. ~\onlinecite{B1}.

We employ the results of Ref. ~\onlinecite{B1} to show that
(details are given in Appendix \ref{SCATFOR}) the current operator
can be written in the form
\begin{align}
\hat{I}(x,t)=\hat{I}^{}_{\rm in}(x,t)-\hat{I}^{}_{\rm out}(x,t)\ ,
\label{HATI}
\end{align}
where $\hat{I}_{\rm in}^{}$ denotes the current operator of
particles moving into the dot,
\begin{align}
\hat{I}^{}_{\rm in}(x,t)&=\frac{e}{h}\int dE dE'e^{i(E-E')t/\hbar}\nonumber\\
\times &a^{\dagger}_{}(E)a^{}_{}(E')e^{i(|k(E')|-|k(E)|)x}\
,\label{IOPIN}
\end{align}
and $\hat{I}^{}_{\rm out}$ is the current operator for particles
moving out of the dot into the reservoir,
\begin{align}
\hat{I}^{}_{\rm out}(x,t)&=\frac{e}{h}\int dE dE'e^{i(E-E')t/\hbar}S^{\ast}(E)S(E')\nonumber\\
\times &a^{\dagger}_{}(E)a^{}_{}(E')e^{i(|k(E)|-|k(E')|)x}\
.\label{IOPOUT}
\end{align}
Here $a(E)$  ($a^{\dagger}(E)$) creates  (destroys) a carrier of
energy $E$ moving into the dot, whose momentum is $k(E)$.

The current operators are used in the Kubo linear-response formula
to compute the (measurable, in principle) currents flowing in the
system. In the present case, the perturbation (resulting from  the
ac voltage source to which the macroscopic gate of Fig. \ref{SYS}
is connected) induces periodic charge accumulation on the dot and
on part of the lead to which it is connected. We assume that the
resulting potential is roughly uniform on the dot and beyond it,
until it drops to zero further away in the lead due to screening,
say at the point $x=-\delta$ (see Fig. \ref{SYS}; the origin of
the $x-$axis is taken at the entrance to the dot). The temporal
derivative of the charge accumulation yields the current at
$x=-\delta$.  Measuring the current at the entrance to the lead
then gives the conductance of the dot in the form
\begin{align}
g(\omega )=\frac{1}{\hbar\omega }\int_{0}^{\infty}dt
e^{i(\omega +i\eta )t}\langle [\hat{I}(0,t ),
\hat{I}(-\delta ,0)]\rangle\ ,\ \ \eta\rightarrow 0^{+}\ .\label{KUBO}
\end{align}
The magnitude of $\delta$, within a rather large range of values,
does not really matter for the total current. Actually, it is argued in Refs.~
\onlinecite{B1} and ~\onlinecite{BUTTI} that for the
relevant (relatively low) frequencies $\omega$, the factor $\exp
[i(k-k')x]$ [see Eqs. (\ref{IOPIN}) and (\ref{IOPOUT})] may be
replaced by unity. Indeed, the expression (\ref{Free-G}) is
obtained upon  setting $\delta$ of Eq. (\ref{KUBO}) equal to zero.
However, in order to establish the different temporal behaviors of
the incoming and the outgoing currents it is useful to retain
$\delta\neq 0$ for the time-being.

Evidently,  inserting Eq. (\ref{HATI}) for the
decomposition of the current operator into the Kubo formula
(\ref{KUBO}), the conductance $g(\omega )$ is decomposed as well
into
\begin{align}
g=g_{\rm in,in}+g_{\rm out, out}- g_{\rm in,out}-g_{\rm out, in}\
,\label{G4}
\end{align}
where $g_{\rm a,b}$ stands for the response of the current flowing
in the a$-$direction to the current (due to the perturbation)
along the b$-$one. We show in Appendix \ref{SCATFOR}
that  two response functions
out of the four listed in Eq. (\ref{G4}) vanish, $g_{\rm out, out}(\omega )
=g_{\rm in,out}(\omega )=0$, and hence there is no response
to the perturbing outgoing current (since $\delta $
is chosen as positive, see Eqs. (\ref{DECOMG})
and the following discussion).
The remaining response functions are finite.
We show in Appendix \ref{SCATFOR}
that
$g_{\rm in ,in}(\omega )=e^{2}/h$, leading to an {\em instantaneous}
response of the incoming current to the perturbing incoming current,
and consequently to the first term in Eq. (\ref{IT}). On the other hand,
$g_{\rm out ,in}(\omega )=(e^{2}/h)\exp [i\omega\tau ]$
[see Eq. (\ref{FG})], causing a {\em delayed} outgoing current.

\section{Capacitance oscillations}
\label{CB}

Monitoring the frequency dependence of the mesoscopic conductance
enables one to study the `quantum capacitance', \cite{GL,B1,BS}
in particular its dependence on the transmission of the quantum dot and on the Fermi energy.
As mentioned above, for `open dots',
previous theoretical studies
of the dependence of Coulomb oscillations on the dot-wire coupling appear to be in a certain conflict \cite{MT,BS}.
In this section we attempt to
shed some light on this intriguing issue.

For noninteracting electrons, the capacitance of the quantum
dot [see Eq. (\ref{QUANT})]
is given by the Wigner-Smith time, i.e., by the energy-derivative of
the transmission phase at the Fermi level.
A possible route to include the effect of interactions on the dot is hence to find
the modifications they cause in the scattering matrix, Eq. (\ref{S-Phase}).
Such a procedure, adopted in Ref. ~\onlinecite{BS},
necessitates the reduction of the (interacting) Hamiltonian to a quadratic form,
e.g., by using an approximation to treat the interactions (see below).
However, once the Hamiltonian is reduced to a quadratic form, one may circumvent
the calculation of the scattering matrix
by using the Friedel sum-rule \cite{HEUSON}
which relates the scattering-matrix  phase to the total occupancy
of the dot levels, $N_{d}$.
As shown in Ref. ~\onlinecite{BLY},
\begin{align}
\frac{d\phi(E_{F})}{dE} = 2 \pi \frac{d N_{d}(E_f)}{dE}\ ,
\end{align}
and therefore the capacitance is given by the
energy-derivative of this total
occupancy.

The Hamiltonian of our model system (see Fig. \ref{SYS})
is written as
\begin{align}
{\cal H}={\cal H}_{\rm wire}^{}+{\cal H}_{\rm dot}^{}+
{\cal H}_{\rm tun}^{}\ .\label{MODELH}
\end{align}
We describe the single-channel wire by
a one-dimensional tight-binding Hamiltonian,
\begin{align}
{\cal H}_{\rm wire}^{}=-J\sum_{\ell =-\infty}^{-1}\Bigl (c^{\dagger}_{\ell}c^{}_{\ell -1}+{\rm hc}\Bigr )\ .\label{WIRE}
\end{align}
Here, $c^{}_{\ell}$ ($c^{\dagger}_{\ell}$)
destroys (creates) an electron on the $\ell$th `site' of the
wire (the lattice constant of the entire system is taken as unity),
and $J$ is the hopping amplitude (in energy units) between adjacent
sites.
The Hamiltonian of the dot, ${\cal H}_{\rm dot}$,
includes the single-particle part and the interactions. \cite{COM5}
The former
is a `continuation' of the single-channel wire,
while the interactions are described by the charging Hamiltonian,
\begin{align}
{\cal H}^{}_{\rm dot}=-J\sum_{\ell =1}^{n_{d}-1}
\Bigl (c^{\dagger}_{\ell}c^{}_{\ell -1}+{\rm hc}\Bigr )+\frac{E_{c}}{2}
\Bigl (\hat{N}-\frac{n_{d}}{2}\Bigr )^{2}\ ,\label{HD}
\end{align}
where $E_{c}$ is the charging energy,  $\hat{N}$
is the number operator on the dot,
\begin{align}
\hat{N}=\sum_{\ell =0}^{n_{d}-1}c^{\dagger}_{\ell}c^{}_{\ell}\ ,
\end{align}
and $n_{d}$ denotes the number of sites on the dot.
Finally, the tunneling Hamiltonian of Eq. (\ref{MODELH})
gives the coupling between the dot and the wire,
\begin{align}
{\cal H}^{}_{\rm tun}=-J_{0}^{}
\Bigl (c^{\dagger}_{0}c^{}_{-1}+c^{\dagger}_{-1}c^{}_{0}\Bigr )\ .
\label{TUN}
\end{align}
The magnitude of the capacitance oscillations is
determined by  the strength of the
coupling between the dot and the wire.

The model Hamiltonian (\ref{MODELH})  is close in spirit
to the system considered by Matveev, \cite{MT}
who has employed advanced techniques to treat the
effects of the electronic correlations. Here we
confine ourselves to the simpler
type of approximations used in the previous
studies of the ac conductance \cite{BS}.
However, since one of our aims in this section is to
explore the relation between the
results of  B\"{u}ttiker and Nigg \cite{BS}
and those of Ref. ~\onlinecite{MT}, it is useful to
show the equivalence of the  Hamiltonian  (\ref{MODELH})
and the Hamiltonian used in Ref. ~\onlinecite{BS}.

To this end, one  diagonalizes  the single-particle
part of the dot Hamiltonian Eq. (\ref{HD}), by introducing
the unitary transformation
\begin{align}
d^{\dagger}_{n}=\sqrt{\frac{n_{d}}{2}}\sum_{\ell =0}^{n_{d}-1}
\sin\Bigl (\frac{\pi (\ell +1)}{n_{d}+1}\Bigr )c^{\dagger}_{\ell}\ ,\label{UNIT}
\end{align}
which turns the dot Hamiltonian into the form
\begin{align}
{\cal H}^{}_{\rm dot}=\sum_{n=1}^{n_{d}}E^{}_{n}d^{\dagger}_{n}d^{}_{n}+\frac{E_{c}}{2}
\Bigl (\hat{N}-\frac{n_{d}}{2}\Bigr )^{2}\ ,\label{HD1}
\end{align}
where the number operator is
$\hat{N}=\sum_{n=1}^{n_{d}}d^{\dagger}_{n}d^{}_{n}$ and
the energy levels are given by
\begin{align}
E_{n}=-2J\cos(\pi n /(n_{d}+1))\ .\label{EN}
\end{align}
In addition, the unitary transformation (\ref{UNIT}) changes
the tunneling part of the Hamiltonian into
\begin{align}
{\cal H}^{}_{\rm tun}&=\sum_{n=1}^{n_{d}}V_{n}^{}\Bigl (
c^{\dagger}_{-1}d^{}_{n}+{\rm hc}\Bigr )\ ,\nonumber\\
V^{}_{n}&=-\sqrt{\frac{2}{n_{d}}}J_{0}\sin \Bigl (
\frac{\pi n}{n_{d}+1}\Bigr )\ ,
\end{align}
such that in the transformed system each level on
the dot is coupled to the wire.

The energy levels on the dot, $E_{n}$, are almost equally-spaced
around mid-band (zero energy in our scheme)
when the dot is rather large, $n_{d}\gg 1$. Equation
(\ref{EN}) then can be approximated by
\begin{align}
E_{n}\simeq n {\rm d}\ ,\ \ {\rm d} =J\frac{2\pi}{n_{d}}\ , \label{SPEC-TRUNC}
\end{align}
where ${\rm d}$ denotes the level spacing. In practice, we allow
$n$ to be within the range $|n|\leq N$ (i.e., $N \ll n_d$ or equivalently $Nd << J$)
thereby truncating the spectrum to $2N+1$ levels. These levels are
also approximately uniformly coupled to the wire, with coupling
energy
\begin{align}
V=-J_{0}\sqrt{\frac{{\rm d}}{J\pi}}\ .
\end{align}
In this way, our model Hamiltonian (\ref{MODELH}) becomes
identical to the one employed in Ref. ~\onlinecite{BS}.

In order to find the capacitance of the dot described by
the truncated Hamiltonian, we need to find the quantum average of the
occupations of the dot levels. This is accomplished by treating
the interactions in the Hartree-Fock approximation, replacing the interaction
term of the dot Hamiltonian ${\cal H}^{}_{\rm d}$ by
\begin{align}
{\cal H}^{}_{\rm HF}=E_{c}\sum_{m,n}\Bigl (Q_{nn}^{}
d^{\dagger}_{m}d^{}_{m}-Q^{}_{nm}
d^{\dagger}_{m}d^{}_{n}\Bigr )\ ,
\end{align}
where the generalized occupancies, $Q^{}_{nm}$,
\begin{align}
Q^{}_{nm}=\langle d^{\dagger}_{n}d^{}_{m}\rangle\ ,
\end{align}
are computed {\em self-consistently}.
Thus, the Hamiltonian of the dot becomes
\begin{align}
{\cal H}^{}_{\rm dot}=\sum_{n} n {\rm d}d^{\dagger}_{n}d^{}_{n}+{\cal H}^{}_{\rm HF}\ .
\end{align}
The occupations of the dot levels
are given by the diagonal ($n=m$) generalized occupancies, but their
determination requires the computation of the non-diagonal ones as well.

The generalized occupancies can be found from the imaginary part of
the dot Green function, in which the coupling to the wire appears as
the width of the dot levels
\begin{align}
Q_{nm}^{}=\int_{-\infty}^{E_F}\frac{dE}{\pi}{\rm Im}
\Bigl [{\cal H}_{\rm dot}^{}-E-i\gamma\Bigr ]^{-1}_{nm}\  . \label{SC-Eq}
\end{align}
The form of the width $\gamma$ is a somewhat subtle point. In our
simple model, the dot is entirely open  when $J=J_{0}$
[see Eqs. (\ref{WIRE}) and (\ref{TUN})],
and then $\gamma$ becomes of the order of the level spacing, ${\rm d}$.
However, it is more realistic to imagine a point contact
coupling the dot and the wire, and then
the transmission between the dot and the wire
depends on the energy of the moving electrons.
We hence follow Ref. ~\onlinecite{BS}
and adopt the result of Brouwer and Beenakker \cite{BB},
which relates the level widths of a very large dot
to the transmission coefficient ${\cal T}$ of the point contact,
\begin{align}
\gamma({\cal T})=\frac{{\rm d}}{\pi{\cal T}}\Bigl (1-\sqrt{1-{\cal T}}\Bigr )^{2}\ .
\label{BEEN}
\end{align}
The transmission coefficient of a point contact modeled by  a saddle point potential
varies rather sharply with the Fermi energy of the reservoir. We have chosen the parameters
of this potential so that
\begin{align}
{\cal T}(E_{F})=\frac{1}{1+e^{-2E_{F}/{\rm d}}}\ . \label{T-OF-EF}
\end{align}
Note that the dot is open when $E_{F}$ exceeds vastly the
level spacing,
and then the width $\gamma $ tends to its maximal value (of the
order of the level-spacing).  When $-E_F$ vastly exceeds ${\rm d}$,
the point contact closes, and concomitantly
the width $\gamma$ vanishes.

The computations described below have been carried out for a dot
with 61 levels ($N=30$) in the truncated Hamiltonian. The Fermi
energy $E_{F}$, the charging energy $E_{c}$, and the width
$\gamma$ [see Eq. (\ref{BEEN})] are measured in units of the
level-spacing ${\rm d}$, with $-2\leq (E_{F}/{\rm d})\leq 5$,
$\gamma$ changing accordingly following Eq. (\ref{T-OF-EF}), and
$E_{c}$ taking various values, indicated on the figures.

A simple iteration algorithm has been used to solve Eq. (\ref{SC-Eq}) for the
generalized occupancies. In order to achieve reliable convergence, Eq.
(\ref{SC-Eq}) has been solved gradually: instead of directly solving with the
chosen parameter set (using some arbitrary initial values), we have solved
with a series of sets, beginning with a certain trivial one (e.g., for
$E_{c}=0$  or for a very weak coupling) and varying  the parameters gradually
till the desired set is reached. For each set, the iteration algorithm has
been initiated with the solution of the previously solved set.

\begin{figure}[h]
\begin{center}
\includegraphics[width=90mm, height=60mm]{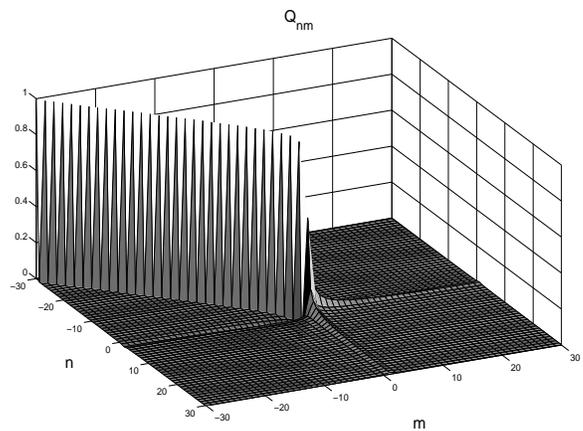}
\end{center}
\caption{Generalized occupancies, $Q_{nm}$, of an open dot with
$E_c={\rm d}$. The Fermi energy is chosen such that level number 3
is half full ($\langle d^{\dagger}_{3}d^{}_{3} \rangle = 0.5$) and
therefore the off-diagonal occupancies are maximal. Still, the
largest ones are smaller than $0.15$.} \label{fig:occ}
\end{figure}

Our computation and the one presented in Ref. ~\onlinecite{BS}
differ in their treatment of the generalized occupancies.
Whereas we keep the non-diagonal occupancies, those have been discarded
by B\"{u}ttiker and Nigg.\cite{BS}
These non-diagonal occupancies vanish
when the dot is isolated. It is therefore
not surprising that our results coincide with those of
Ref. ~\onlinecite{BS} when the dot is weakly coupled to the lead.
However, when the coupling is strong and the dot is `open',
these non-diagonal occupancies, albeit being rather small
(see Fig. \ref{fig:occ})
play a crucial role.

\begin{figure}[t]
\includegraphics[width=9cm]{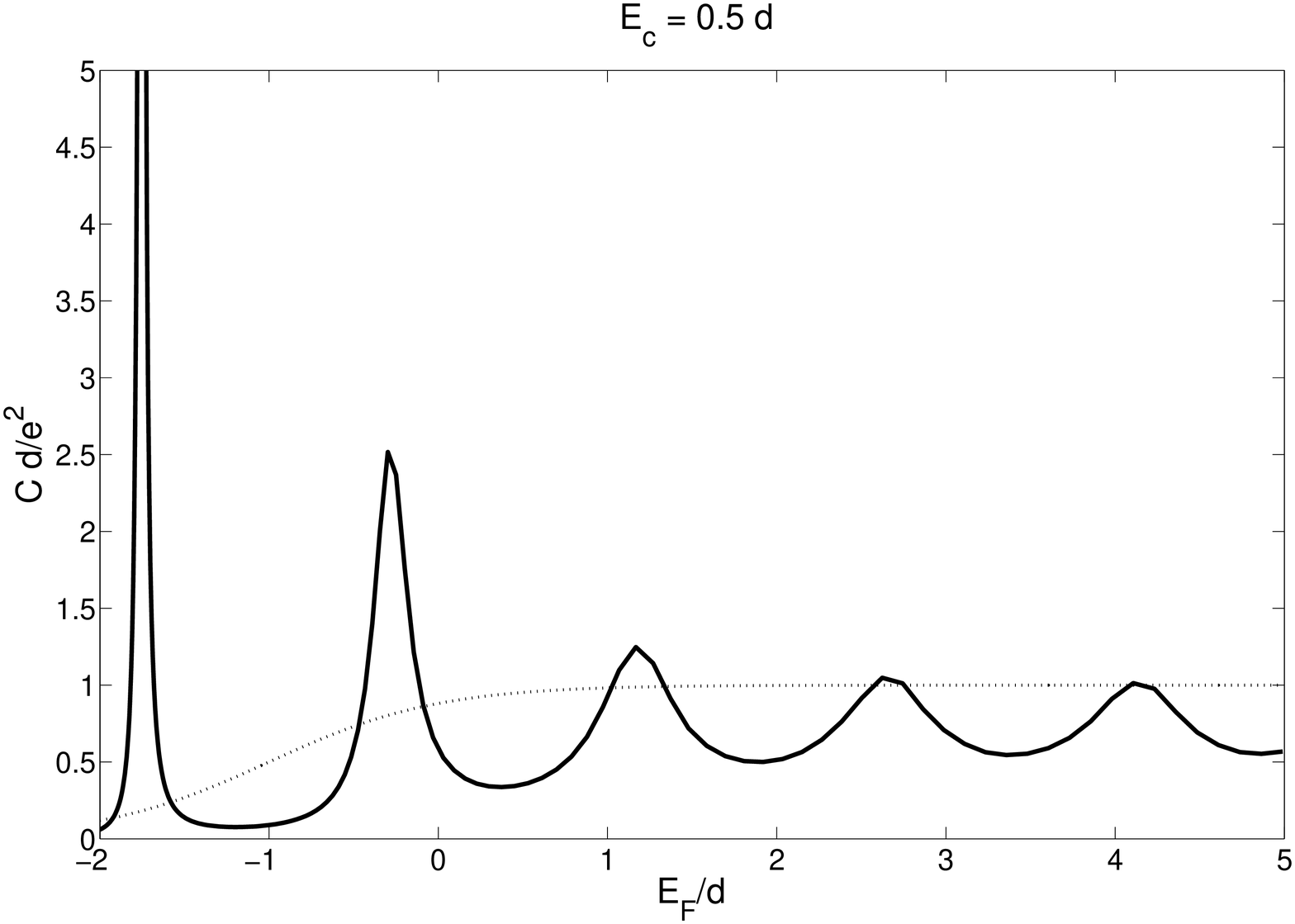}
\includegraphics[width=9cm]{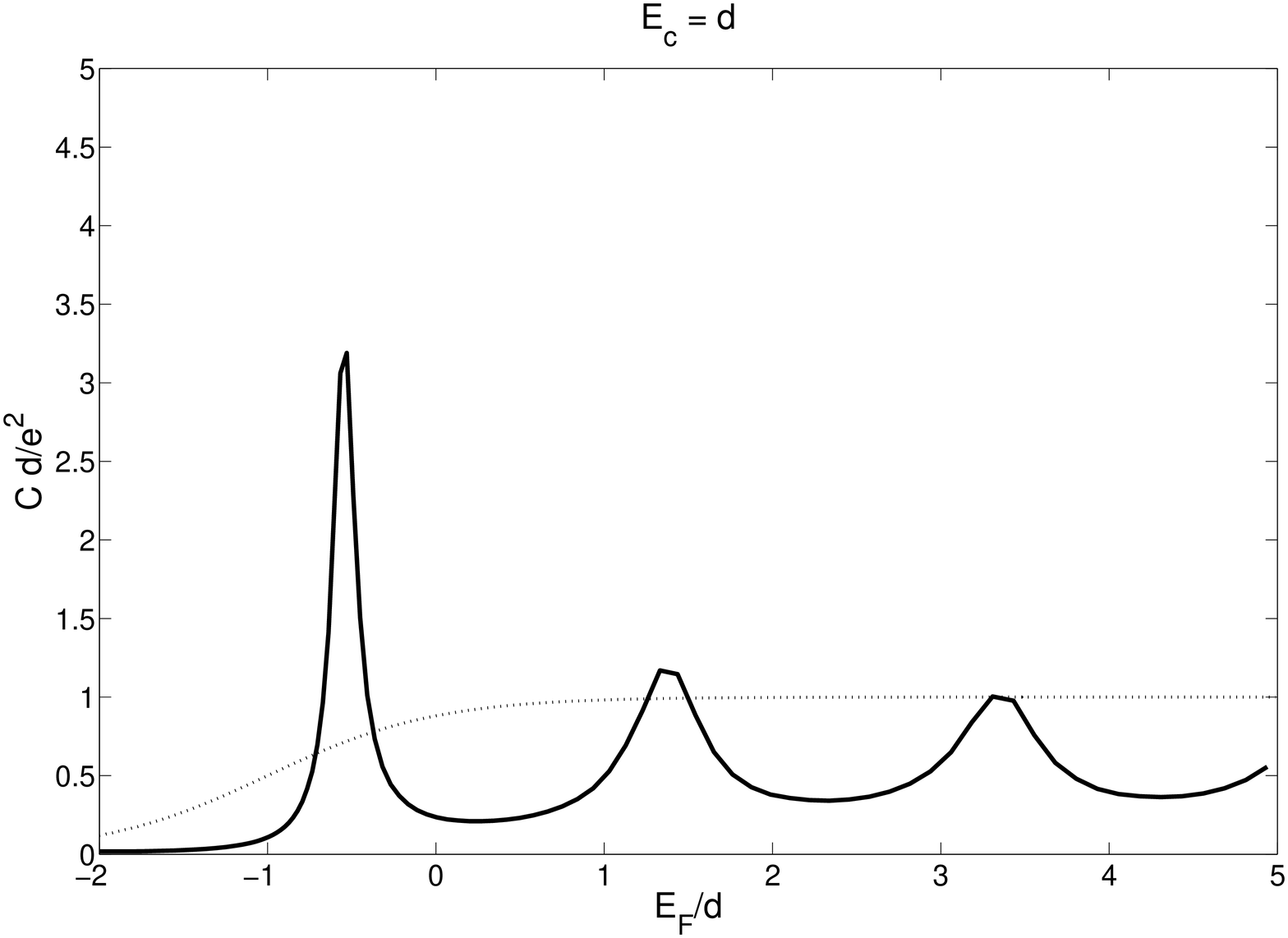}
\caption{The capacitance as a function of the
Fermi energy computed without including the
non-diagonal occupancies, for two values of $E_c/{\rm d}$ (solid lines).
The dashed lines are the
transmission at the entrance of the dot computed from Eq. (\ref{T-OF-EF}).
The capacitance is measured in units of $e^2/{\rm d}$.} \label{fig:Hartree}
\end{figure}

\begin{figure}[t]
\includegraphics[width=9cm]{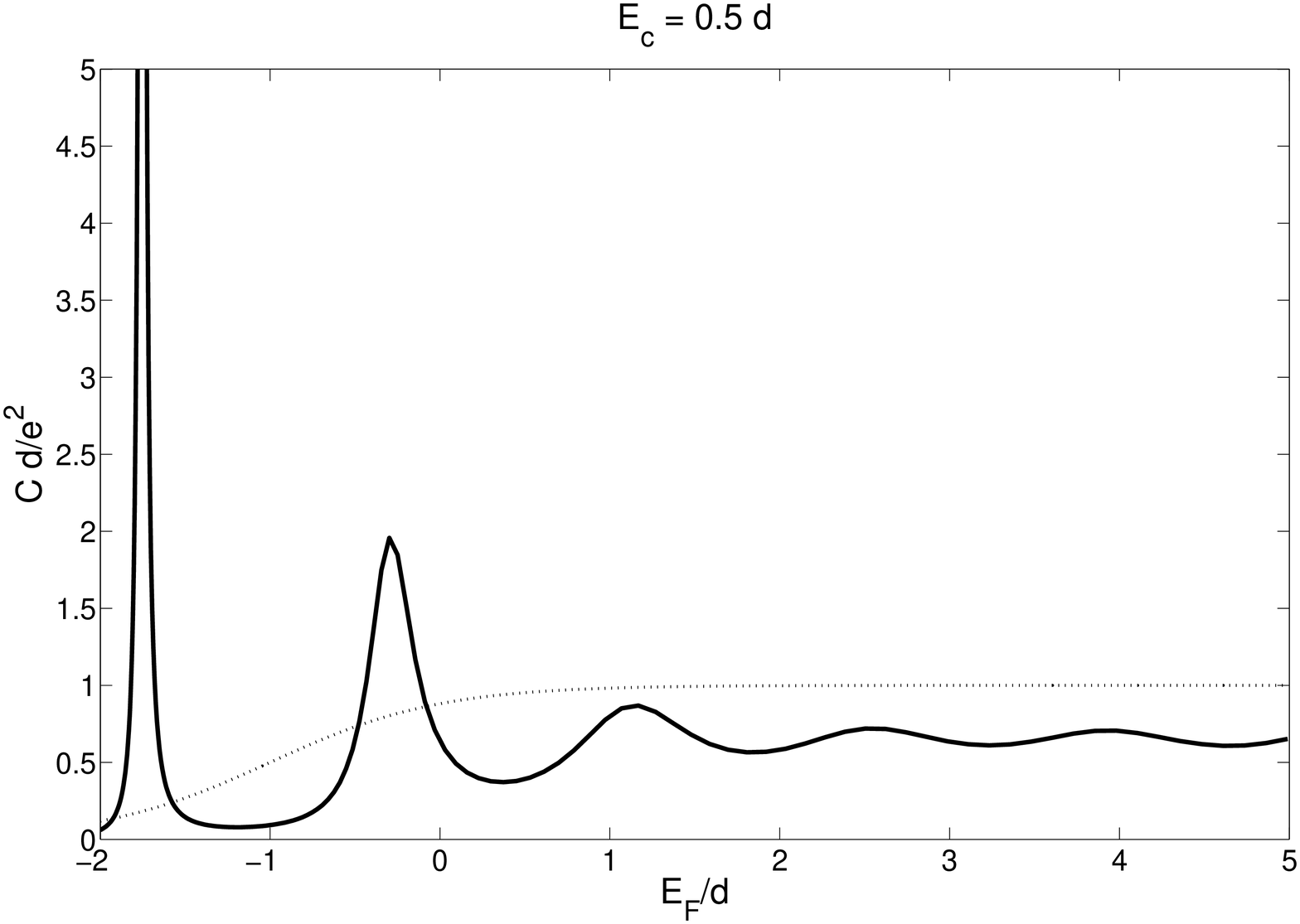}
\includegraphics[width=9cm,height=5.8cm]{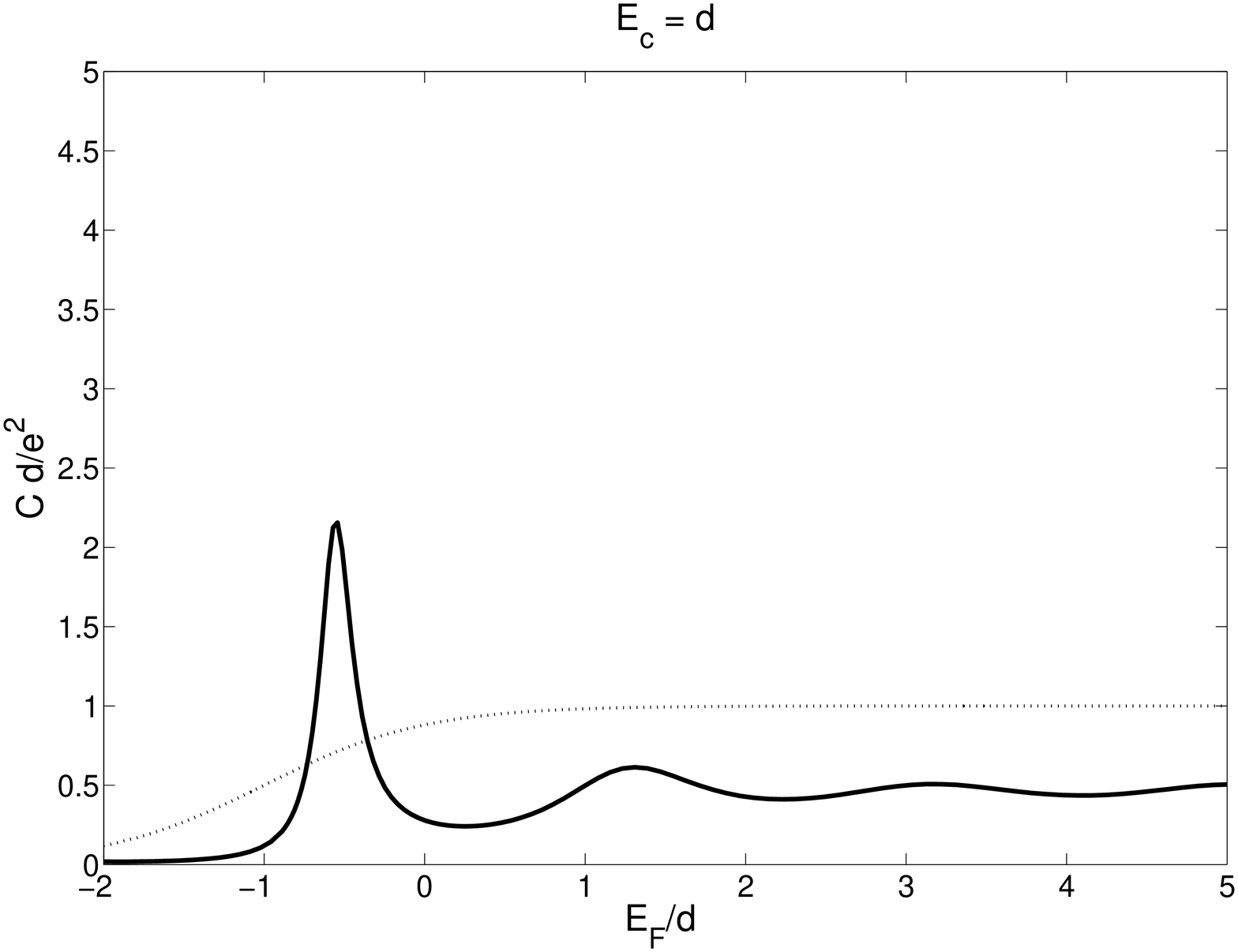}
\caption{The capacitance as a function of the
Fermi energy computed with the
non-diagonal occupancies, for two values of $E_c/{\rm d}$ (solid lines).
The dashed lines are the
transmission at the entrance of the dot computed from Eq. (\ref{T-OF-EF}).
The capacitance is measured in units of $e^2/{\rm d}$.
} \label{fig:HartreeFock}
\end{figure}

A detailed comparison between our results and those
obtained according to the approximation carried out in
Ref. ~\onlinecite{BS} is shown in Figs.
\ref{fig:Hartree} and
\ref{fig:HartreeFock}. Those figures depict the capacitance
as function of the Fermi energy (solid lines). Figure
\ref{fig:Hartree} exhibits the capacitance  computed {\em without}
the non-diagonal occupancies, i.e., according to the procedure of
Ref. ~\onlinecite{BS}, while Fig.
\ref{fig:HartreeFock} shows the same quantity computed in the presence of
the non-diagonal occupancies, i.e., employing the full
Hartree-Fock approximation
of the model. In all panels, the dotted lines represent
the transmission between the dot and lead.
At small values of the Fermi energy, i.e., on the
left part of each plot, the dot is loosely
coupled to the lead [see Eq. (\ref{T-OF-EF})], and the
Coulomb-blockade peaks are well-defined, separated from
each other  by
${\rm d}+ E_c$. As the transmission increases, i.e.,
on the right side of each plot,
the capacitance oscillations are more smeared.
However, the smearing is much more
pronounced when all the Hartree-Fock
terms are included in the computations: whereas in
Fig. \ref{fig:Hartree} the peaks persist and
their heights even increase with interaction strength,
they decay into
a weaker oscillation which does not increase with the  interaction strength in
Fig. \ref{fig:HartreeFock}.


\begin{figure}[h]
\centering
\includegraphics[width=9cm]{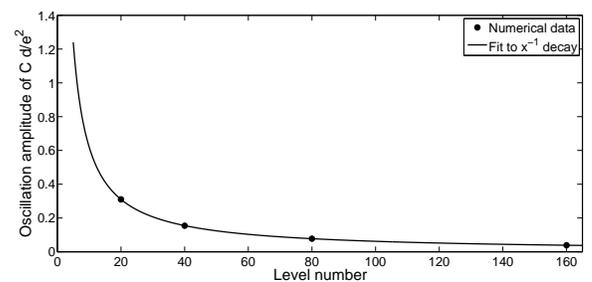}
\caption{The peak height of the interaction-free capacitance C,
measured in units of $e^2/{\rm d}$, as function of the level
number ($2N+1$). \label{fig:oscdecay}}
\end{figure}

The weaker oscillations appearing on the right
side of each of the plots in Fig. \ref{fig:HartreeFock}
are not a signature of the CB effect but rather
an artifact related to the truncation of the spectrum.
Note that these oscillations appear already in the
interaction-free case ($E_c=0$) [see Fig.
\ref{fig:HartreeFock}
] and decay rather than increase as the interaction strengthens.
Figure \ref{fig:oscdecay} presents the decay of
these oscillations as a function of the dot-level number, $N$.
Thus, {\em at least up to order $1/N$, the Hartree-Fock approximation
establishes the vanishing of Coulomb blockade effects in an open dot,
even for small values of $E_{c}/{\rm d}$}.

\begin{figure}
\includegraphics[width=9cm]{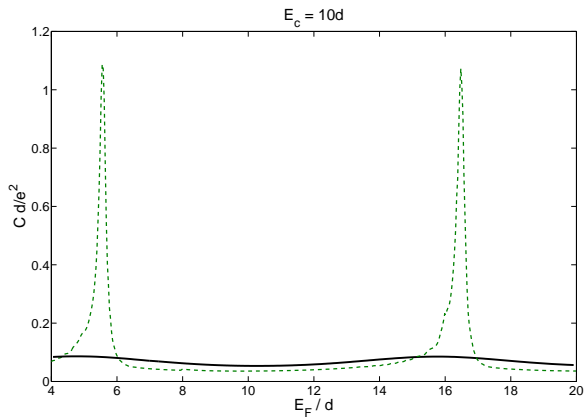}
\caption{The capacitance as a function of the
Fermi energy computed with and without
off-diagonal occupancies (solid and dashed lines,
respectively), for $E_c= 10{\rm d}$. The transmission
throughout the scanned energy range is essentially
one, following Eq. (\ref{T-OF-EF}). The capacitance
is measured in units of $e^2/{\rm d}$.} \label{fig:HartreeBig}
\end{figure}

Using a different approach, Matveev \cite{MT} predicted the vanishing of CB
effects in open quantum dots, for large values of $E_{c}$/d. Figure
\ref{fig:HartreeBig} portrays the capacitance for $E_c/{\rm d}= 10$ , both
for the case where the non-diagonal occupancies are included (solid line) and
for the case when they are absent (dashed line). Note the disappearance of
the CB peaks in the first case, compared to the second one.

\section{Conclusions}

We have considered here the ac conductance of a quantum dot connected by an ideal
single-channel wire to an electron reservoir and
driven by a nearby gate at
frequency $\omega$. This conductance can, at low
frequencies, be represented in
terms of a conventional RC circuit. It was predicted
in Ref. ~\onlinecite{B1} and confirmed
experimentally in Ref. ~\onlinecite{GL} that R is given
by half the quantum resistance and
that $C$ exhibits, in general, quantum oscillations.
These are due to the
varying density of states of the dot.

In Sec. \ref{DC} we showed that the results for
the ac conductance could be
interpreted in terms of an ``incoming" current, synchronous with the driving
voltage and an ``outgoing" current delayed by $\tau$ with respect to the latter,
[see Eq. (\ref{IT})] where $\tau$ is the Wigner-Smith
delay time of the dot, given by the
derivative of the reflection phase of the dot with
respect to the
incoming energy. The identification of these two currents
is obtained using a space-dependent linear response analysis.
Equations (\ref{QUANT}) follow almost
trivially, and provides a vivid
physical picture for the results.

In Sec. \ref{CB} we address the issue of
Coulomb blockade oscillations in the limit in which the dot is
strongly coupled to the lead (dot-level width comparable with its
level spacing). Reference ~\onlinecite{BS} found, from a modified
Hartree procedure, that the oscillations persist. On the other hand,
it had been found by Matveev \cite{MT} that they should vanish in the limit
of a fully open dot.  We have treated the problem
including the full Hartree-Fock approximation, i.e., taking into account the non-diagonal occupancies. We have found that
the CB oscillations do vanish in the limit of a open dot, up to possible $1/N$-type corrections, where N is the number of the levels on the dot. this result follows from a Hartree-Fock treatment, and does not require the more elaborate treatment of the electronic correlations on the dot \cite{MT}.
\pagebreak

\begin{acknowledgements}
We wish to thank Markus B\"{u}ttiker for instructive discussions and Simon E.
Nigg for a helpful correspondence. This work was supported by the German
Federal Ministry of Education and Research (BMBF) within the framework of the
German-Israeli project cooperation (DIP), the Israel Science Foundation (ISF)
and by the Converging Technolgies Program of the Israel Science Foundation
(ISF), grant No 1783/07.
\end{acknowledgements}

\appendix*
\section{The current operators in the scattering formalism}
\label{SCATFOR}

In the scattering-formalism approach one expands the current
operator, $\hat{I}$, in terms of operators which create (destroy)
the incoming or the outgoing electrons \cite{COM3},
\begin{align}
\hat{I}(x,t)&=\sum_{\sigma\sigma '}\int dE dE'e^{i(E-E')t/\hbar}\nonumber\\
&\times a^{\dagger}_{\sigma}(E)a^{}_{\sigma '}(E')I_{\sigma\sigma
'}(x;E,E')\ . \label{CUROP}
\end{align}
Here, $a^{\dagger}_{\sigma}(E)$   [$a^{}_{\sigma}(E)$] creates
(destroys) a carrier of energy $E$ moving in the direction defined
by $\sigma$: $\sigma =+$ denotes incoming (into the dot)
particles, whose momentum is positive, while $\sigma =-$ denotes
the particles going away from the dot.  These operators obey the
anti-commutation relations
\begin{align}
[a^{\dagger}_{\sigma}(E),a^{}_{\sigma '}(E')]^{}_{+}=
\delta_{\sigma\sigma '}\delta (E-E')\ ,
\end{align}
and are normalized such that
\begin{align}
\langle a^{\dagger}_{\sigma}(E)a^{}_{\sigma '}(E')\rangle =
\delta_{\sigma\sigma '}\delta (E-E') f(E)\ ,\label{AVER}
\end{align}
where $f(E) $ is the Fermi function of the particles in the
reservoir. The two sets of operators, those belonging to the
incoming particles and those of the outgoing ones are related by
the scattering matrix,
\begin{align}
a^{}_{-}(E)=S(E)a^{}_{+}(E)\ .\label{PSM}
\end{align}
The matrix elements of the quantum-mechanical current density
operator in the scattering states which appear in Eq.
(\ref{CUROP}) are
\begin{align}
I_{\sigma\sigma '}(x;E,E')=\frac{e}{h}\frac{v_{k}+
v_{k'}}{2\sqrt{|v_{k}v_{k'}|}} e^{i(k' -k)x}\ ,\label{IX}
\end{align}
where $v_{k}$ is the velocity of the $k-$ state, $v_{k}=\partial
E/(\hbar\partial k)$, and $k=\sigma |k|\equiv k(E)$ (and
similarly, $k'=\sigma '|k'|\equiv k(E')$).

At low enough frequencies, the relevant energies in Eq.
(\ref{CUROP}) are close to the Fermi level. Then, the magnitude of
both $v_{k}$ and $v_{k'}$ is $v_{F}$, the Fermi velocity, and they
only differ in their signs. \cite{COM4} It follows that the velocity factor of
Eq. (\ref{IX}) becomes $\sigma\delta_{\sigma\sigma'}$, i.e.,
\begin{align}
I_{\sigma\sigma '}(x;E,E')=\frac{e}{h}\delta_{\sigma\sigma '}
\sigma e^{i(k'-k)x}\ .\label{APR}
\end{align}
As a result one may unambiguously distinguish in the current
operator Eq. (\ref{CUROP}) the incoming current operator,
$\hat{I}^{}_{\rm in}$, for which both $\sigma $ and $\sigma '$ are
positive, from the outgoing one, $\hat{I}^{}_{\rm out}$, for which
$\sigma=\sigma '=-$, leading to Eqs. (\ref{IOPIN}) and
(\ref{IOPOUT}). For brevity, we have omitted in the main text the
subscript $+$ from the operators $a$ and $a^{\dagger}$.

We next consider the partial response functions
that appear in Eq. (\ref{G4}).
Each of those involves the same thermal average
[see Eqs. (\ref{IOPIN}), (\ref{IOPOUT}), and (\ref{KUBO})],
\begin{align}
\langle
&[a^{\dagger}_{}(E)a_{}^{}(E'),a^{\dagger}_{}
(\tilde{E})a^{}_{}(\tilde{E}')]\rangle\nonumber\\
 =&\delta (E-\tilde{E}')\delta
(E'-\tilde{E})(f(E)-f(E'))\ ,
\end{align}
where $a^{}(E)\equiv a^{}_{+}(E)$, and
$a^{\dagger}(E)\equiv a^{\dagger}_{+}(E)$.
Performing also the temporal integration, we find
\begin{widetext}
\begin{align}
g^{}_{\rm in , in}(\omega )&=i\frac{e^{2}}{h^{2}\omega} \int
dEdE' \frac{f(E)-f(E')}{E-E'+\omega +i\eta} e^{i(|k(E')|-|k(E)|)\delta}\ ,\nonumber\\
g^{}_{\rm out , out}(\omega )&=i\frac{e^{2}}{h^{2}\omega} \int
dEdE' \frac{f(E)-f(E')}{E-E'+\omega +i\eta}
 e^{i(|k(E)|-|k(E')|)\delta}\ ,\nonumber\\
g^{}_{\rm in, out}(\omega )&=i\frac{e^{2}}{h^{2}\omega} \int
dEdE' \frac{f(E)-f(E')}{E-E'+\omega +i\eta}
e^{i(|k(E)|-|k(E')|)\delta}S(E)S^{\ast}(E')\ ,\nonumber\\
g^{}_{\rm out , in}(\omega )&=i\frac{e^{2}}{h^{2}\omega}\int
dEdE' \frac{f(E)-f(E')}{E-E'+\omega +i\eta}
 e^{i(|k(E')|-|k(E)|)\delta}S^{\ast}(E)S(E')\
.\label{MG}
\end{align}
Here  $|k(E)|$ stands for $+\sqrt {2m E}/\hbar $.
In each of the expressions of Eqs. (\ref{MG}) we may carry out one
of the energy integrations. For example, the integrals appearing in
$g^{}_{\rm in , in}(\omega )$ may be written as
\begin{align}
&i\int dEdE' \frac{f(E)-f(E')}{E-E'+\omega +i\eta}
e^{i(|k(E')|-|k(E)|)\delta} =i\int dE f(E)\int dE'\Biggl
(\frac{e^{i(|k(E')|-|k(E)|)\delta}}{E-E'+\omega
+i\eta}-\frac{e^{i(|k(E)|-|k(E')|)\delta}}{E'-E+\omega
+i\eta}\Biggr )\ ,\label{EXAM}
\end{align}
\end{widetext}
and then the $E'-$integration is performed by closing the
integration contour in the appropriate half of the complex plane.
In the right-hand-side of Eq. (\ref{EXAM}), we close
the contour of the first term in the upper half-plane,
and that of the second term in lower half-plane.
Since $\exp [i|k(E')|\delta ]$ is just the scattering matrix of an
ideal piece of wire (of length $\delta$), it must be analytic
in the upper half of the complex $E'-$plane and decay to zero
at very large values of $E'$
(and similarly  $\exp [-i|k(E')|\delta ]$
is analytic in lower half-plane). Computing
then the $E'-$integration  by the Cauchy theorem yields
\begin{align}
&i\int dEdE' \frac{f(E)-f(E')}{E-E'+\hbar\omega +i\eta}
e^{i(|k(E')|-|k(E)|)\delta}\nonumber\\
 &=2\pi\int dE \Bigl (f(E)-f(E+\hbar\omega )\Bigr )
e^{i(|k(E+\hbar\omega)|-|k(E)|)\delta}\nonumber\\
&
\simeq h \omega\ ,
\end{align}
where in the last step we have taken the very low-temperature
limit, and have used $\omega \delta /2v_{F}\ll 1$.
The same procedure, when applied to
$g^{}_{\rm out , out}(\omega )$ yields zero,
since for this quantity the appropriate closures of
the integrations are the same as for the ones in Eq. (\ref{EXAM}),
but now the contours do not enclose any poles.
The choice of the contours for the integrations in
$g^{}_{\rm in, out}(\omega )$ and in $g^{}_{\rm out, in}(\omega )$
is opposite to the one of Eq. (\ref{EXAM}).
For this reason
$g^{}_{\rm in, out}(\omega )$ vanishes while
$g^{}_{\rm out, in}(\omega )$
does not (note that the combination
 $\exp [i|k(E')|\delta ]S(E')$
is again a scattering matrix, and as such is analytic
in the upper half-plane).
It therefore follows that
\begin{align}
&g_{\rm in , out}^{}(\omega )=g_{\rm out ,out}^{}(\omega )=0\
,\nonumber\\
&g_{\rm in ,in}(\omega )=\frac{e^{2}}{h} \
,\nonumber\\
&g_{\rm out ,in}^{}(\omega
)=\frac{e^{2}}{h}\int_{E_F-\hbar\omega}^{E_F}\frac{dE}{\hbar\omega
}S^{\ast}(E)S(E+\hbar\omega )\ .\label{DECOMG}
\end{align}
Namely, the current responds to the perturbing incoming current
alone; had we interchanged the locations of the perturbing and the
responding currents, i.e., had we chosen $\delta <0$, we would
have obtained the complementary result, for which ``in" is
interchanged with ``out" in Eqs. (\ref{DECOMG}).
As the relevant energies in the final integral yielding the expression for
$g_{\rm out ,in}^{}(\omega
)$ are confined to the vicinity of the Fermi level, it follows that [upon using Eq.
(\ref{S-Phase})]
\begin{align}
g_{\rm out ,in}^{}(\omega
)=\frac{e^{2}}{h}e^{i\omega \tau}\ ,\label{FG}
\end{align}
where $\tau$ is given by Eq. (\ref{WST}).


\begin{thebibliography}{999}
\bibitem{COM2}

It is straightdorward to verify that corretions to this value (i.e.,
derivatives of the Wigner-Smith time) do not affect the low-frequency
properties of the conductance.
In fact, the first derivative is cancelled in the expansion of $g(\omega )$
up to second order in the frequency.
\bibitem{COM5}

Our discussion is confined to the case where interactions
are dominant on the dot, while they are assumed to be screened
away on the wire (see Fig. \ref{SYS}).
\bibitem{COM3}

In detailing the linear-response derivation within scattering
theory, we confine ourselves to the simple case at hand:  a dot
connected by a single-channel wire  to a single reservoir.
\bibitem{COM4}

One may wonder whether a full treatment of the electronic
dispersion (beyond the linearization around the Fermi level) may
modify our picture of incoming and outgoing current operators.
However, it turns out that such deviations affect the ac conductance
only at frequency orders higher than the second.
\end{thebibliography}
\end{document}